\documentclass[prl,twocolumn,showpacs,amsmath,amssymb,superscriptaddress]{revtex4}

\usepackage{graphicx}
\usepackage{dcolumn}
\usepackage{bm}

\begin{document}


\title{Manipulating the torsion of molecules by strong laser pulses}

\author{C.B. Madsen}
\affiliation{Lundbeck Foundation Theoretical Center for Quantum
System Research, Department of Physics and Astronomy, Aarhus University, 8000 Aarhus C, Denmark}

\author{L.B. Madsen}
\email[Corresponding author: ]{bojer@phys.au.dk}
\affiliation{Lundbeck Foundation Theoretical Center for Quantum
System Research, Department of Physics and Astronomy, Aarhus University, 8000 Aarhus C, Denmark}

\author{S.S. Viftrup}
\affiliation{Department of Chemistry, Aarhus University, 8000
Aarhus C, Denmark}

\author{M. P. Johansson}
\affiliation{Department of Chemistry,  Aarhus University, 8000
Aarhus C, Denmark}

\author{T.B. Poulsen}
\affiliation{Department of Chemistry, Aarhus University, 8000
Aarhus C, Denmark}

\author{L. Holmegaard}
\affiliation{Department of Chemistry, Aarhus University, 8000
Aarhus C, Denmark}

\author{V. Kumarappan}
\affiliation{Department of Chemistry, Aarhus University, 8000
Aarhus C, Denmark}

\author{K.A. J{\o}rgensen}
\affiliation{Department of Chemistry, Aarhus University, 8000
Aarhus C, Denmark}

\author{H. Stapelfeldt}
\email[Corresponding author: ]{henriks@chem.au.dk}
\affiliation{Department of Chemistry and Interdisciplinary
Nanoscience Center (iNANO), Aarhus University, 8000 Aarhus C,
Denmark}


\date{\today}

\begin{abstract}
A proof-of-principle experiment is reported, where torsional motion
of a molecule, consisting of a pair of phenyl rings, is induced by
strong laser pulses. A nanosecond laser pulse spatially aligns the
carbon-carbon bond axis, connecting the two phenyl rings, allowing a
perpendicularly polarized, intense femtosecond pulse to initiate
torsional motion accompanied by an overall rotation about the fixed
axis. The induced motion is monitored by femtosecond time-resolved
Coulomb explosion imaging. Our theoretical analysis accounts for and
generalizes the experimental findings.
\end{abstract}

\pacs{33.15.Hp, 33.80.Rv, 42.50.Hz}

\maketitle

A non-resonant laser field applies forces and torques on molecules
due to the interaction between the induced dipole moment and the
laser field itself. If the field is intense, but non-ionizing, the
forces and torques can be sufficient to effectively manipulate the
external degrees of freedom of isolated gas phase molecules. In
particular, the intensity gradient of a focused laser beam may
deflect~\cite{stapelfeldt:1997}, focus~\cite{chung:jcp:2001} and
slow~\cite{Fulton:NP:2006} molecules through the dependence of the
non-resonant polarizability interaction on the intensity. Likewise,
the dependence of the induced dipole interaction on molecular
orientation has proven highly useful for controlling the alignment
and rotation of a variety of
molecules~\cite{stapelfeldt:2003:rmp,villeneuve:prl:2000}. Molecular
manipulation by induced dipole forces extends beyond the external
degrees of freedom and can also be applied to the internal degrees
of freedom such as vibrational motion~\cite{Niikura:prl:2003}.
Notably,  the electrical field from laser pulses can modify energy
potential barriers such that photoinduced bond breakage of a small
linear molecule is guided to yield a desired final
product~\cite{Sussman:science:2006,Sussman:pra:2006}.

\begin{figure}
    \centering
   \includegraphics[width=1\columnwidth]{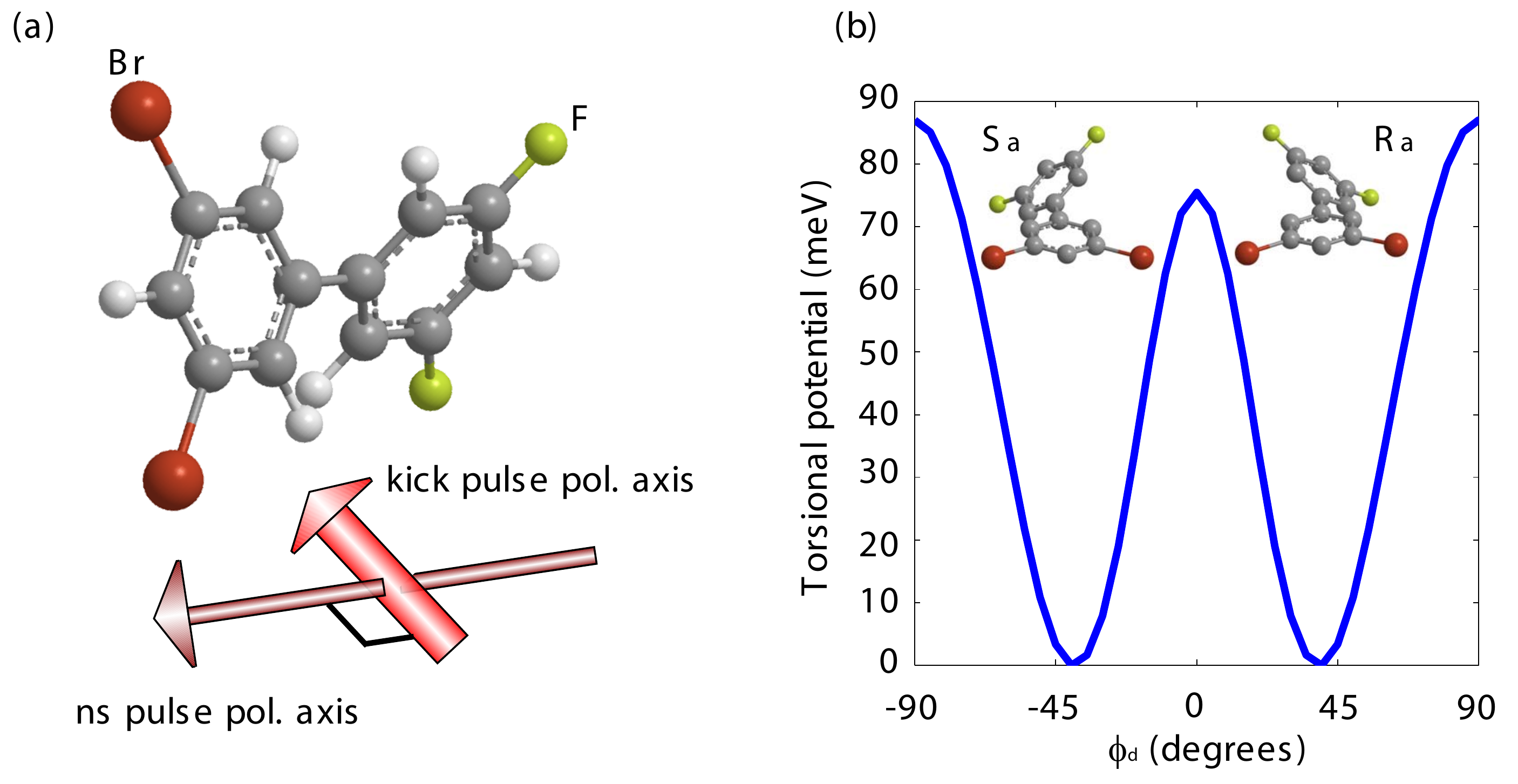}
 \caption{
 (Color online)  (a) Model of DFDBrBPh and the laser geometry of the experiment. The molecule consists of a pair of phenyl rings. The substituted Br and F atoms are needed to discriminate the two rings in the experiment.
(b) The torsional potential of DFDBrBPh~\cite{mikael08}. The torsion
is quantified by the dihedral angle, $\phi_d$, between the two
phenyl rings. The minima at dihedral angles of $\phi_d=\pm39^\circ$
result in the R$_a$ and S$_a$ enantimomers (see text).}
  \label{fig1}
\end{figure}

Here, we extend the use of the non-resonant polarizability
interaction to achieve a transient modification of the torsional
potential of the molecule 3,5-diflouro-3',5'-dibromo-biphenyl
(DFDBrBPh), thereby inducing torsional motion of the two phenyl
rings (Fig.~\ref{fig1}). The results are accounted for by our
theoretical analysis, and we point at two very diverse aspects of
the current work: (i) The potential application of twisted molecules
in, e.g., molecular junctions, where they may serve as ultrafast
(picosecond) molecular
switches~\cite{ratner:science:2003,tour:science:2001,chen:cp:2002,Seideman:prl:2007}.
(ii) DFDBrBPh has two conformations that are mirror images of each
other, the R$_a$ and S$_a$ enantiomers~\cite{Eliel}. An extension of
the present experiment will provide a unique temporally resolved
study of the important chemical process
de-racemization~\cite{Faber:2001,Fujimura:cpl:1999,Shapiro:prl:2000,kroner:cp:2007},
where one enantiomer is selectively converted into the other.

The experiment is carried out on isolated molecules at rotational
temperatures of a few Kelvin. Under these conditions each molecule
is initially localized in the $-39^\circ$ or 39$^\circ$ conformation
(Fig.~1b) and no thermally induced transitions between the two
occur. The scheme to obtain torsional motion is based on the
combination of two laser beams (Fig.~\ref{fig1}a). First, a $1064$
nm, 9 ns (FWHM) linearly polarized laser pulse of peak intensity
$7\times10^{11}$ W/cm$^2$ fixes the carbon-carbon (C-C) axis between
the two phenyl rings in the laboratory frame by adiabatic
alignment~\cite{stapelfeldt:2003:rmp,kumarappan:2006:jcp}. This
pulse is intense enough to keep the axis tightly confined, yet weak
enough to modify the torsional potential only slightly. Next, an
$800$ nm laser pulse of intensity $5\times10^{12}$ W$/$cm$^2$  and
duration (FWHM) $700$ fs, which we will refer to as the kick pulse,
is applied with its polarization perpendicular to the aligned C-C
bond axis. This polarization geometry ensures that the kick pulse
primarily influences torsional motion while avoiding excitation of
other normal modes. At time $t_p$ with respect to the kick pulse, a
linearly polarized, intense $800$ nm, $25$ fs (FWHM),
$2\times10^{14}$ W/cm$^2$ pulse appears and removes several
electrons from the molecules, thereby triggering Coulomb explosion
into ionic fragments. In particular, the Br$^+$ and F$^+$ fragment
ions recoil in the planes defined by the two phenyl rings. By
recording the velocities of both ion species with two-dimensional
ion imaging~\cite{kumarappan:2006:jcp}, we thus determine the
instantaneous orientation of each of the two phenyl rings at $t_p$.
The time-resolved ion images are displayed in Fig.~2a.
\begin{figure}
    \centering
    \includegraphics[width = 1.0\columnwidth]{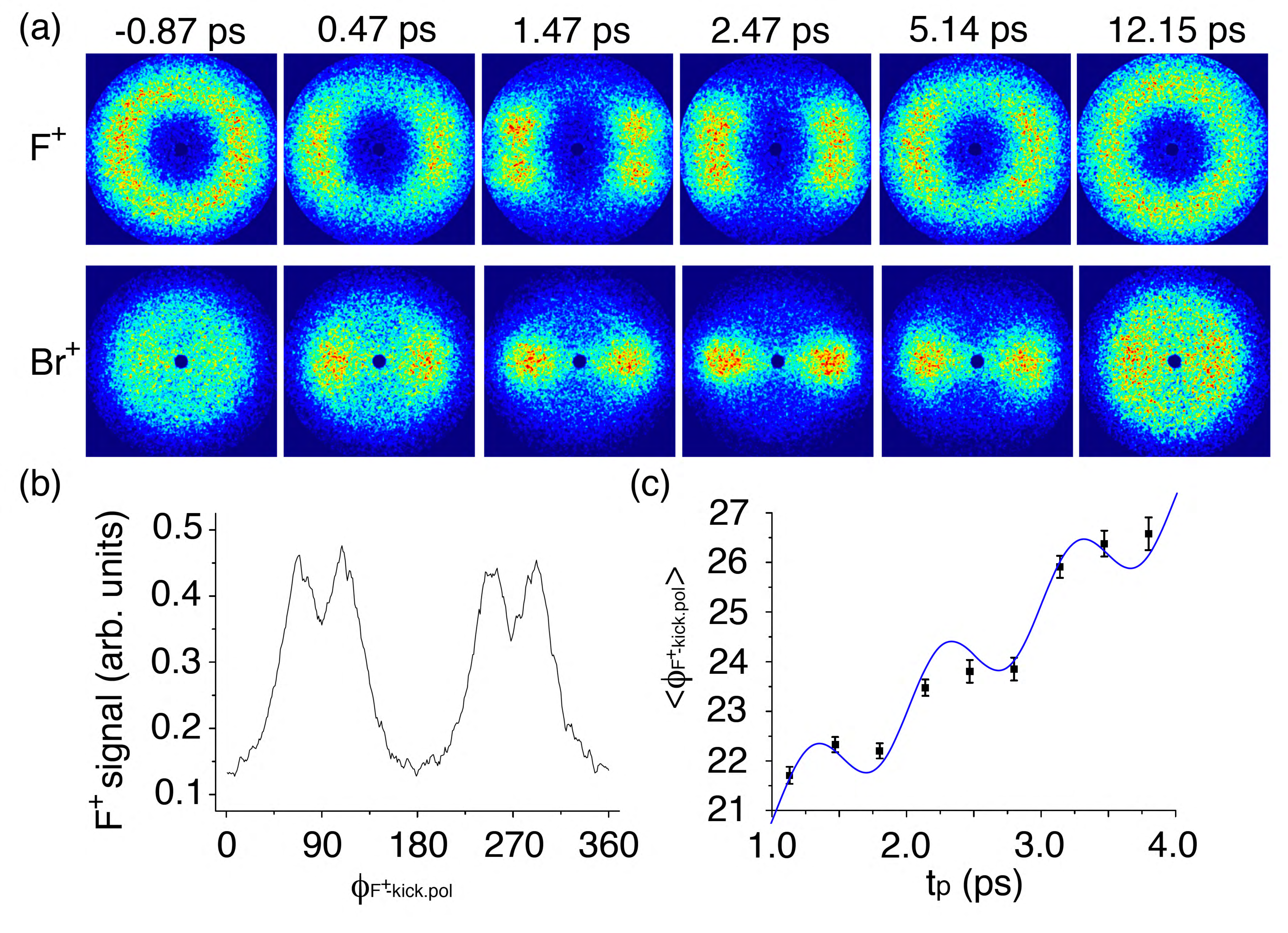}
 \caption{
 (Color online)
 (a)  Ion images of F$^+$ and Br$^+$ fragments at probe times $t_p$.
 The ns pulse is polarized perpendicularly to the image (detector)
plane and the  $5\times10^{12}$ W$/$cm$^2$,
 $700$ fs (FWHM) kick pulse is polarized horizontally. (b) Angular
distribution of the F$^+$ ions, at $t_p$~=1.47 ps, obtained by
radially integrating the corresponding F$^+$ ion image between the
F$^+$ ion recoil and the kick pulse polarization for fixed angle,
$\phi_{\text F^\text+ \text{- kick. pol}}$. The splitting of the
pairwise peaks is twice the the average angle, $\langle\phi_{\text
F^\text+ \text{- kick. pol}}\rangle$, between the F$^+$ ion recoil
and the kick pulse polarization. (c) $\langle\phi_{\text F^\text+
\text{- kick. pol}}\rangle$ as a function of $t_p$, for times where
a clear four-peak structure is visible in the angular distributions.
The curve is a fit of the sum of a linear and a harmonic function to
the experimental points (squares). }
 \label{fig2}
\end{figure}

To establish that the ns pulse aligns the C-C axis, we have applied
the probe pulse at $t_p=-0.87$ ps. This results in almost circularly
symmetric ion images (Fig.~\ref{fig2}a) and the small deviation from
circular symmetry is explained by noting that the kick pulse has a
finite value at $-0.87$ ps. The absence of ions in the innermost
region, most clearly seen in the F$^+$ image, shows that the C-C
bond is aligned perpendicular to the detector plane and that the
rotation of the molecule around this axis is uniform. At $t_p =
0.47$ ps the deviation from circular symmetry is very clear and the
ions start to localize around the polarization direction of the kick
pulse. The F$^+$ ions remain radially confined away from the center,
which shows that the kick pulse does not perturb the alignment of
the C-C bond axis. Rather, it initiates an overall rotation of the
molecule around this axis as is expected since the torque imparted
by the kick pulse forces the second most polarizable axis (SMPA),
perpendicular to the C-C axis and located 11$^\circ$ away from the
Br-phenyl ring, to align along the kick pulse polarization on a time
scale determined by the kick strength~\cite{Viftrup:prl:2007}. At
$t_p = 1.47$ ps the Br$^+$ ions are localized around the
polarization axis and the F$^+$ ion distribution exhibits a
four-peaked structure. This behavior is compatible with alignment of
the SMPA along the kick pulse polarization. In practice and
consistent with theory (Fig.~3a), the SMPA alignment is not strong
enough to resolve the two Br$^+$ signals located at $\pm 11^\circ$
with respect to the SMPA. It is, however, sharp enough to resolve
the two pairs of F$^+$ ion signals due to the larger offset
(28$^\circ$) of the F-phenyl rings from the SMPA. The further
localization of the Br$^+$ signal at 2.47 ps shows that the
Br-phenyl planes have rotated into stronger alignment with the kick
pulse polarization. Had the dihedral angle remained unchanged, the
F$^+$ ion image should exhibit a distinct four-dot structure similar
to the image at 1.47 ps but with a larger angular splitting. The
four-dot structure at 2.47 ps is, however, significantly blurred
compared to the case at 1.47 ps. Thus, we conclude that the kick
pulse sets the molecule into rotation around the C-C axis and
initiates torsional motion. At later times, the ion signals
gradually broaden due to continued overall rotation around the C-C
axis with dihedral dynamics imposed.

Further insight into the effect of the kick pulse is obtained by
plotting the average angle between the F-phenyl rings and the kick
pulse polarization as a function of $t_p$ (Fig.~2c). The increase
from 22.5$^\circ$ at 1.47 ps to 26.5$^\circ$ at 3.8 ps shows that
the F-phenyl rings gradually move away from the kick pulse
polarization due to the overall rotation of the molecule around the
C-C axis. The concurrent oscillations show that the overall rotation
is accompanied by a periodically varying motion in $\phi_d$. We
estimate the period to be $\sim 1$ ps and the amplitude to $\sim
0.6^\circ$ for this oscillation.

\begin{figure}
    \centering
    \includegraphics[width=1.0\columnwidth]{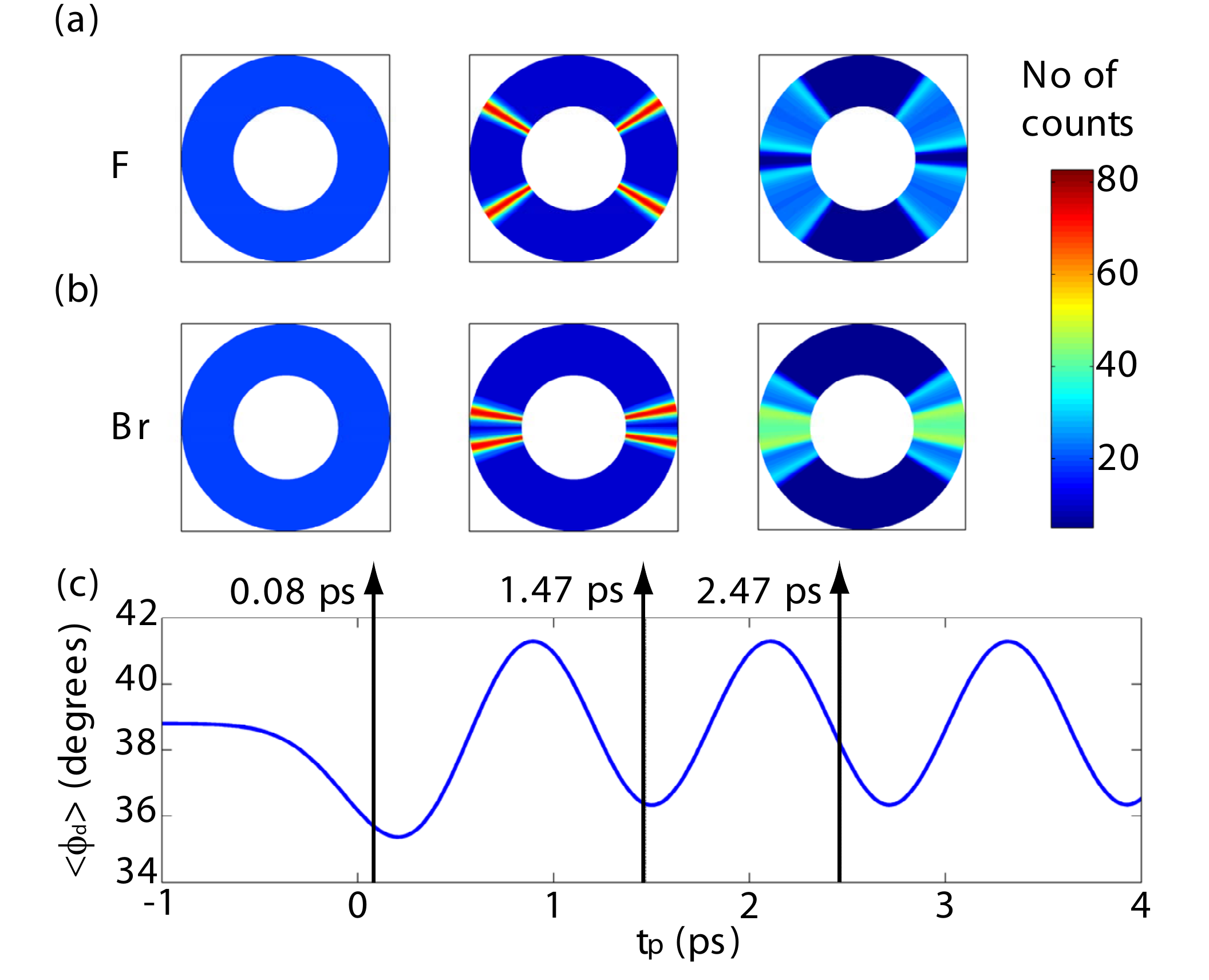}
 \caption{(Color online) Angular distributions of (a) F-phenyl and (b) Br-phenyl rings at $t_p=0.08$, 1.47 and 2.47 ps. (c) Expectation value of the dihedral angle for a
molecule starting out with the SMPA aligned along the kick pulse
polarization. The kick pulse is as in Fig.~2.
 }
  \label{fig3}
\end{figure}

We now present our physical model substantiating the experimental
findings. In agreement with the observations, we assume that the C-C
axis is perfectly aligned. Also, we neglect all normal modes, except
the lowest one, corresponding to torsion. Within these
approximations the task is reduced to describing the coupled
rotations of the two phenyl rings. The experimental observables are
the angles $\phi_\text{Br}$ and $\phi_\text{F}$ of the rings with
respect to the kick pulse polarization axis. In the theoretical
treatment it is, however, convenient to refer to the dihedral angle
$\phi_d = \phi_\text{Br} - \phi_\text{F}$ and the overall rotation
around the C-C axis described by the weighted azimuthal angle $\Phi
= (1- \eta) \phi_\text{Br} + \eta \phi_F$, with $\eta$ given in
terms of the moments of inertia, $\eta =
I_\text{F}/(I_\text{Br}+I_\text{F})$. This change of coordinates
separates the dynamics of the molecule into two rotations of
different time scales. The first, $\phi_d$, is an internal rotation
corresponding to torsional motion. The time scale of torsion is ps,
as can be inferred from the torsional potential. The second
rotation, $\Phi$, is an overall rotation with a period of ns, as
follows from the total moment of inertia $I=I_\text{Br}
+I_\text{F}$. This separation motivates the introduction of a
semi-classical model, where $\phi_d$ is treated fully quantum
mechanically, while $\Phi$ is treated classically. Briefly, due to
the kick pulse the field-free torsional state of energy $E_\nu$
evolves as [we apply atomic units, $m_e=e=a_0=\hbar=1$]
$\vert\chi_\nu\rangle\to\vert\chi_\nu^\Phi(t)\rangle=\sum_{\nu'}c_{\nu'}^\Phi(t)e^{-iE_{\nu'}(t-t_0)}\vert\chi_{\nu'}\rangle$,
with $t_0$ a time prior to the kick pulse. The time-dependent
coefficients satisfy the differential equations
$\dot{c}_{\nu'}^\Phi(t)=-i\sum_\nu
c_\nu^\Phi(t)e^{-i(E_\nu-E_{\nu'})(t-t_0)}\langle\chi_{\nu'}\vert
V_\text{kick}(\Phi,t)\vert\chi_\nu\rangle$, with the kick pulse
polarization interaction
\begin{align}
V_\text{kick}(\Phi,\phi_d,t)&=-\frac{1}{4}F_0^2(t)[\alpha_\text{xx}(\phi_d)\cos^2(\Phi+\eta\phi_d)\nonumber\\
&+\alpha_\text{yy}(\phi_d)\sin^2(\Phi+\eta\phi_d)\label{eq:kickpot}\\
&-2\alpha_\text{xy}(\phi_d)\cos(\Phi+\eta\phi_d)\sin(\Phi+\eta\phi_d)].\nonumber
\end{align}
Here $\alpha_{ij}$'s are dynamic polarizability tensor components
obtained by our quantum chemistry calculations and $F_0$ is the kick
pulse field envelope. The effect of the kick pulse on the overall
rotation amounts to an angular momentum kick and consequently
\begin{equation}\label{eq:Phit}
\Phi(t)=\left.\Phi_0-t\frac{1}{I}\left(\frac{\partial}{\partial\Phi}\int_{-\infty}^\infty
dt' \langle
V_\text{kick}(\Phi,t')\rangle\right)\right\arrowvert_{\Phi=\Phi_0},
\end{equation}
where $\langle\rangle$ denotes averaging over the ensemble of
torsional states.

Figure 3 shows the results of a calculation with laser parameters
identical to the experimental values and an initial rotational
temperature of 0 K. Prior to the kick pulse the angular
distributions of the Br- and F-phenyl rings (left panels, Figs.~3a
and~3b) are isotropic as in the experiment. Maximum alignment of the
SMPA occurs at 1.3 ps and the confinement of the F-phenyl rings at a
large angle with respect to the kick pulse polarization (cf.~middle
panel, Fig.~3a) explains the distinct four-dot structure observed at
$t_p = 1.47$ ps (Fig.~2a) in the experimental F$^+$ ion image. Also,
at $t_p = 1.47$ ps the confinement of the Br-phenyl rings at a small
angle with respect to the kick pulse polarization predicts a much
less distinct, if any, four-dot structure in good agreement with the
Br$^+$ ion image.  At $t_p = 2.47$ ps the angular localization of
the F-phenyl rings has broadened (right panel, Fig.~3a) and a
blurred four-dot structure is seen, consistent with the experimental
result at $t_p = 2.47$ ps. The distribution of the Br-phenyl rings
is also broadened (right panel, Fig.~3b) but remains localized
around the kick pulse polarization fully consistent with the Br$^+$
ion distribution, recorded at 2.47 ps.

The theoretical value $\langle \phi_d \rangle$ exhibits oscillations
(Fig.~3c) with a period of $\sim1.2$ ps and amplitude of $\sim
2.45^\circ$. The period agrees well with the experimental value
($\sim 1$ ps), and smaller modulation in $\langle\phi_d\rangle$ is
expected in the experiment ($\sim 0.6^\circ$) since here the SMPA is
not pre-aligned. The behavior is ascribed to a wave packet of
vibrational modes in the torsional double well potential
(Fig.~\ref{fig1}b) for a molecule starting out with the SMPA
aligned. The qualitative agreement of Figs.~2b and 3c verifies the
interpretation of the kick pulse inducing time-dependent torsional
motion.

Now we point at some possible future applications of induced
torsional motion. For one thing, DFDBrBPh represents a class of
molecules, where the conductivity can be controlled by manipulating
the torsion. Such a molecule used as a molecular junction between
two conductors thus offers an attractive alternative to mechanical
break junctions~\cite{Wu}. In particular, the fact that the dihedral
angle with the present laser pulse-based method can be altered on a
ps time scale opens intriguing possibilities for studying ultrafast
modulation and switching of electrical charge flow.

Another new application within reach is a time-resolved study of
de-racemization~\cite{Faber:2001,Fujimura:cpl:1999,Shapiro:prl:2000,kroner:cp:2007},
where one enantiomer is steered into its mirror form. To this end,
we break the inversion symmetry with respect to the C-C bond axis by
orienting each molecule rather than just aligning in order to
discriminate between the two enantiomeric forms.
Theory~\cite{friedrich:1999:jpca} and
experiment~\cite{tanji:pra:2005} show that orientation can be added
to 3D alignment by combining the ns alignment pulse with a static
electric field. Next, to reduce the role of overall rotation, the
SMPA can be aligned prior to the kick pulse by employing an
elliptically rather than a linearly polarized ns pulse
\cite{stapelfeldt:2003:rmp,Seideman:prl:2007}. Finally, the
interaction strength between the molecule and the kick pulse needs
to be increased either through higher intensity, a longer kick pulse
or by trains of synchronized kick
pulses~\cite{leibscher:2003:prl,bisgaard:prl:2004,lee:jpb:2004}.
\begin{figure}
    \centering
    \includegraphics[width=0.7\columnwidth]{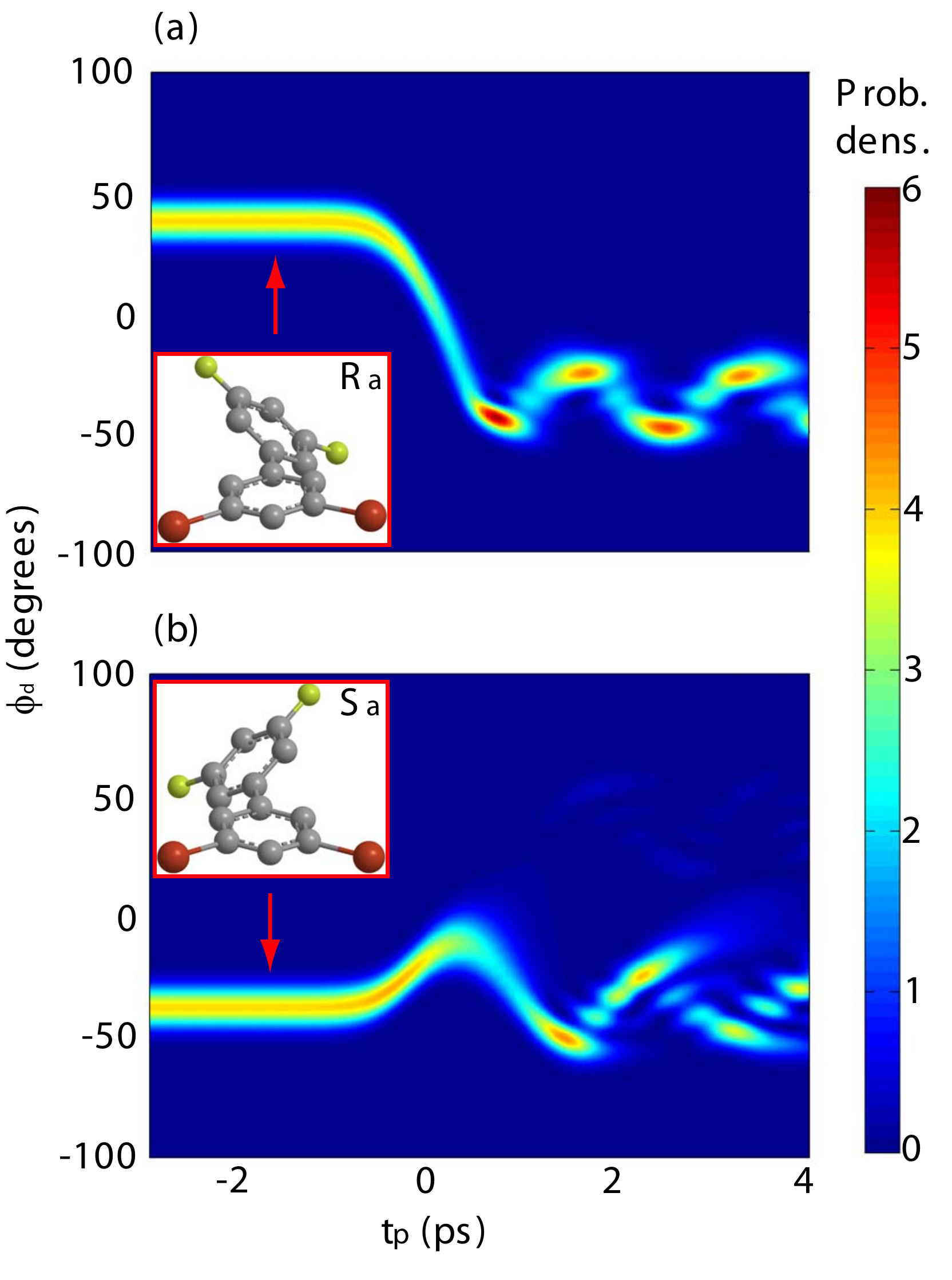}
\caption{ 
(Color online) Time evolution of the dihedral angle for a molecule
starting out as (a) an R$_a$ or  (b) an S$_a$ enantiomer. Initially,
the molecule is 3D oriented with the Br-phenyl end pointing out of
the paper and the SMPA aligned at an angle of 13$^\circ$ with
respect to the kick pulse polarization. The kick pulse triggering
the torsional motion has a peak intensity of $1.2\times10^{13}$
W/cm$^2$ and duration (FWHM) of $1.0$ ps. The torsional barrier is
reduced by 1/4 rather than increasing the kick strength. Such
modification of the torsional potential may be accomplished by
replacing DFDBrBPh with, e.g., halogen substituted
biphenylacetylene.}
  \label{fig4}
\end{figure}
Assuming initial orientation and confinement of the SMPA, we have
calculated the $\phi_d$ dynamics for both conformations of a
molecule closely related to DFDBrBPh. The results are shown in
Fig.~\ref{fig4}, and clearly, the present method would allow for a time-resolved study of the transition from one enantiomer and into the other. A quantitative analysis of
the efficiency of the process shows that after the pulse $99\%$ of
the molecules starting out as R$_a$ changed into S$_a$ enantiomers,
whereas only $13\%$ of the S$_a$ enantiomers changed into R$_a$. The
inverse process causing an excess of R$_a$ enantiomers, is simply
achieved by inverting the orientation of the molecules.

In conclusion, we have performed fs time-resolved studies of
torsional motion by a combination of aligning, kick and ionizing
pulses, and we supported the observations by an accompanying theory.
We believe that the present work opens new directions of
cross-disciplinary research firmly anchored in strong-field physics
as exemplified by ultrafast swithing in molecular junctions and fs
time-resolved studies of de-racemization.

The work was supported by the Danish Research Agency, the Danish
National Research Foundation and the Carlsberg Foundation.


\end{document}